\documentclass{aastex}
\usepackage{spr-astr-addons}

\RequirePackage{color}

\begin{document}

\title{Tkachenko waves, glitches and precession in neutron stars}
\slugcomment{Not to appear in Nonlearned J., 45.}
\shorttitle{Tkachenko waves, glitches and precession}
\shortauthors{Popov}

\author{S.B. Popov\altaffilmark{1}} 
\affil{Sternberg Astronomical Institute\\ 
Russia, 119991, Moscow, Universitetski pr. 13\\
polar@sai.msu.ru}


\begin{abstract}
Here I discuss possible relations between free precession of neutron stars, Tkachenko
waves inside them and glitches.
I note that the proposed precession period of the isolated neutron star 
RX J0720.4-3125  (Haberl et al. 2006) is consistent with the period of Tkachenko waves 
for the spin period $8.4$~s. 
Based on a possible observation  of a glitch in RX J0720.4-3125  
(van Kerkwijk et al. 2007), I propose a simple model, in which long 
period precession is powered by Tkachenko waves generated by a glitch. 
The period of free precession, determined by a NS oblateness,
 should be equal to the standing Tkachenko wave period 
for effective energy transfer from the standing wave to the precession motion.
A similar scenario can be applicable also in the case of the PSR B1828-11.
\end{abstract}



\section{Introduction}

Isolated neutron stars (NSs) being non-spherical bodies are expected to demonstrate free
precession (for a brief review see, for example,  \citealt{l2003}). 
However, examples of this phenomena are less than few, and even in
rare cases when a precession-like behavior is observed different interpretations
can be discussed (even not related to precession, see for example
\cite{rg2006}).  

The problem of long period free precession in NSs is 
a long standing one. A NS can precess if it is non-spherical 
and rotation axis does not coincide with a principal axis. Typically, biaxial objects
are discussed, so deviation from spherical symmetry can be described by one parameter --
oblateness (see \cite{alw2006} for a discussion of triaxial model).
Expected values of NS oblateness (due to rotation or influence of strong magnetic fields) 
can naturally lead to precession periods about one year. 
The precession period is equal to $P_\mathrm{prec}=P/\epsilon$.
Here $P$ is the spin period of a NS, $\epsilon$ -- its oblateness, and
$P_\mathrm{pres}$ -- precession period. Measured precession periods require oblateness 
about $10^{-8}$.
 
However, discussing dynamics of NSs it is necessary to take into accout the network of 
superfluid vortices inside them. The neutron 
superfluid liquid in the interior of a NS participates
in rotation via formation of quantized vortex lines. 
The density of these lines per unit area 
is $n=2\Omega / k$. Here $\Omega=2\pi /P$ is spin frequency, and $k=h/2m_\mathrm{n}$,
where $m_\mathrm{n}$ is a neutron mass (see, for example, \citealt{st1983}, Ch. 10). 
The vortices exist in the core of a NS,
where they can interact with superfluid (superconducting) protons and normal electrons,
and in the crust, where they can pin to it. 

Coupling of superfluid neutron vortices with electrons in a core results in damping of
free precession \citep{ao1987}. But the time scale of this damping is long enough, 
according to these authors.
For spin period about 1 second it is $\sim 400$~--~$10^4$~$P_\mathrm{prec}$ \citep{ao1987}.
Still, this time scale is much shorter than a NS age, so some excitation mechanism
is necessary for precession. As it is discussed below, in the presented 
model excitation is due to 
a glitch.

 A kind of pinning (``immobilization'') of vortices can also happen in the core due 
to interactions with magnetic flux tubes (see discussion, for example, in 
\citealt{l2007}). In this case, the moment of inertia of ``pinned'' neutrons
(which is about $I$)
is about 10 times larger, than the moment of the remaining parts of a NS, $I_\mathrm{c}$.
So, $P_\mathrm{prec}\sim 0.1 P$.

A different kind of problem appears if pinning in the crust is taken into account.
For absolute pinning no long period precession is possible. Instead, 
the period of precession becomes equal to $P (I/I_\mathrm{p})$, here $I$ is NS moment
of inertia, $I_\mathrm{p}$ is the moment of inertia of pinned superfluid in the crust.
Typically it is expected that $I_\mathrm{p}/I \sim 10^{-2}$ \citep{s1977}, 
and the precession period is just $\sim100$~$P$ if the absolute pinning is
valid. However, \cite{ao1987} showed that this is not the case due to finite
temperature. Because of thermal effect always there is vortex creep which
allow the pinned superfluid to follow precession. 

The best example of a NS with precession-like behavior is PSR B1828-11. 
The proposed period is about 511 days with a harmonic at 256 days \citep{sls2000}. 
Most of discussions related to free  precession deal with this source.
In particular, the problem of non-existence of long period precession for strong pinning
is typically confronted with observaions of PSR B1828-11.

Recently, appeared another possible example of long period free precession in NSs.
The existence of $\sim 7$ years precession period 
in one of a small group of 
isolated  NSs (called XDINS -- X-ray Dim Isolated NSs, 
or ICoNS -- Isolated Cooling NSs, or the {\it Magnificent Seven}) --
RX J0720.4-3125 -- was suspected \citep{hetal2006}\footnote{A slightly
different period $\sim$ 4.3 years was proposed by van Kerkwijk and Kaplan
(2007) based on timing analysis. However, these authors consider a model
with a glitch to fit better due to a rapid change in spectral properties,
see van Kerkwijk et al. (2007).}.
So, this object was added to the list, and the paradoxical situation of long precession
in presence of superfluid vortices
 was reconsidered by \cite{l2006,l2007}. 
 This author proposed that either protons 
in NS interiors are type I superconductors, or neutrons in the outer core are normal 
(i.e., not superfluid). More recently \cite{gaj2007} demonstrated that 
for long spin period and small precession angles NSs can have long precession periods
(note, that for PSR B1828-11 the precession angle is proposed to be small,
about few degrees \citep{sls2000},
but for RX J0720.4-3125 it can be larger, $>$~10 degrees \citep{hetal2006}).
So, according to \cite{gaj2007},
the conclusion by \cite{l2006} and other authors that in the strong drag regime 
$P_\mathrm{prec}\sim 0.1 P$  can be under doubt due to a short wavelength  instability.

Clearly, the problem of free precession in NSs is far from being solved completely.
In this brief note, based on coincidence between Tkachenko wave
period and 
precession period in cases of PSR B1828-11 and RX J0720.4-3125, 
I discuss a mechanism to support precession in isolated NSs.

\section{Tkachenko waves}

A simple model for long period precession of isolated NSs proposed 
here is
related to the so-called Tkachenko waves \citep{t1966}. 
These are displacement waves in the vortex line array that exist in rotating superfluid,
or in other words
a kind of sound waves 
propagating in the lattice of neutron votices perpendicular to them.
A good introduction to the Tkachenko waves physics can be found in 
the paper by \cite{ag1982}.

Already in early 70-s this phenomena was suggested to explain periodic
 modulations 
in NSs \citep{r1970, d1971}. At that time motivation had been related to reported wobbling
of the Crab pulsar, which was not confirmed by later observations.
  Then this approach was nearly forgotten, and only recently 
\cite{ns2008} returned to consideration of Tkachenko waves in NSs.
In particular, they demonstrated that behavior of PSR B1828-11 can be explained 
by these waves.

According to \cite{r1970} (see also an example given by \cite{d1971})  
the period of a standing Tkachenko wave in a  NS can be estimated as:

\begin{equation}
P_\mathrm{T}=(2\pi/k) (1/V_\mathrm{T})\sim 1.77 \, R_6 P^{1/2} \, {\rm yrs}. 
\end{equation}
Here $V_\mathrm{T}$ -- wave velocity, which in a simple case depends only on the spin period
($V_\mathrm{T}\sim P^{-1/2}$) and fundamental constants. 
$R_6$ -- the core radius normalized to 
$10^6$~cm 
(with such normalization the equation provides an estimate close to the
upper limit for 
the period). The estimate is made for the mode with $k R =5$.
Spin period of a NS, $P$, is given in seconds.

As one can see, this period is of order of those related to free precession.


\section{Scenario for long period precession}

The proposed precession periods for RX J0720.4-3125 and PSR B1828-11
are very similar to the Tkachenko wave periods for these stars.
 For RX J0720.4-3125 the period of precession is proposed 
to be equal to $\sim 7$ years \citep{hetal2006} or $4.3$ years
\citep{vkk2007}, spin period of this NS is equal to 
$\sim 8.4$~s (see, for example, \cite{h2007} for a review on XDINS). 
For PSR B1828-11  the precession period is equal to $\sim500$~days, 
while the spin period is equal to 0.4~s, for this object the coincidence between
precession and Tkachenko wave periods was already mentioned 
(see, for example, \citealt{gk2004, ns2008}).  
For $R$ about few~--~10 km one  obtains that the precession period is consistent with 
$P_\mathrm{T}$ for both NSs (also the mode can be used as a parameter, 
however everywhere here I use $kR=5$, as proposed by \citealt{r1970}).
 
Moreover, in the case of RX J0720.4-3125 pulse profile modulations and
spectral changes are observed. Period modulations 
related just to the Tkachenko waves alone hardly can be responsible for such evolution. 
Precession is necessary. 
The idea, proposed here is the following: energy stored 
in standing Tkachenko wave can power precession of a NS. The necessary condition 
for effective energy transfer can be the equality of $P_\mathrm{T}$  and free 
precession period. The latter one depends on oblateness of a NS, which can be 
due to strong magnetic fields. The former one depends only on the spin frequency.    
Coincidence between the two characteristic time scales
which depend on different quantities should be not a very frequent occasion.
However, one more condition is necessary -- it is necessary to generate the 
standing wave. The necessity of these two conditions can explain why precessing isolated NSs
are so rare.

\cite{r1970} notes that Tkachenko waves can be generated by starquake glitches. 
Energy of precession is $E_\mathrm{prec}=I\Omega \Omega_\mathrm{prec} \theta^2/2$, 
where $\theta$ is the precession amplitude \citep{j2004}.
For PSR B1828-11 this energy is about $3 \, 10^{36}$~erg.
This value is significantly smaller than typical glitch energy. 
In the case of RX J0720.4-3125 a glitch was proposed by  \cite{vketal2007}. 
The glitch energy was estimated by them as $\sim 10^{37.5}$~erg ($\Delta
\nu/\nu ~\sim 5\, 10^{-8}$).
Of course, it is necessary to say, that the energy of the glitch was
estimated according to the fit with cubic model \citep{vketal2007}, and so
the value can be different if after the glitch the evolution is due to
precession. Still, as an order of magnitude estimate the value from
\cite{vketal2007} can be used.  
\cite{hetal2006} estimated the amplitude of precession $>10^\circ$, 
but warn about uncertainties of their model. The period of precession is about 7 years. 
The energy of the precession motion then appears to be  
$\sim 3 \, 10^{35} \, (\theta/10^\circ)^2$~erg.
With these values in hand  the energy 
of the glitch is enough to drive precession even for large $\theta$
if efficiency of energy transfer is not very low.

\cite{vketal2007} relate a ``jump'' in spectral properties of 
RX J0720.4-3125 to the glitch. In my opinion, this means that  
it is more probable that the glitch was due to a quake, not due to vortex lines 
unpinning (or accretion episode etc., see below).

Taking altogether, 
for RX J0720.4-3125  the following scenario is proposed: 
a glitch (most probably due to a starquake) generates Tkachenko waves; 
the 
period of a standing Tkachenko wave is equal to the 
free precession period
 for this NS; 
due to the standing wave precession starts after a glitch, or just there is
an energy input into the pre-existing precession motion.

 Tkachenko waves periodically change the spin frequency and moment of inertia of a NS.
Waves move perpendicular to the vortex lines, which are parallel to the spin axis.
The moment of inertia of a star can be non-symmetric respect to this axis, for example
if oblateness is due to strong magnetic field. I speculate that periodic modulation of 
spin frequency and all components of moment of inertia in resonance with the 
precession period (determined by oblateness) would lead to energy transfer
from Tkachenko waves to the precession motion.


\section{Discussion}

Absence of free precession in absolute majority of isolated NSs indicates
that this phenomena needs some rare coincidence in properties of a NS.
Here it is proposed that it is necessary to have:
\begin{itemize}
\item $P_\mathrm{T}\approx P_\mathrm{prec}$,  
\item a glitch to generate Tkachenko waves.
\end{itemize}

Instead of the proposal by \cite{l2007} -- {\it "A slowly-precessing neutron star 
cannot glitch"} -- I propose another: {\it slow precession is powered by glitches via 
Tkachenko waves}. 

Observations  of RX J0720.4-3125 are roughly consistent with this scenario.
On the other hand, in the case of PSR B1828-11 no glitches have been observed. 
However, it is necessary to
study for how long precession can survive after a glitch. 
If an old estimate by \cite{ao1987}, $400$~--~$10^4\, P_\mathrm{prec}$ is valid, 
then this time is long enough. If precession is periodically excited by glitches
via Tkachenko waves even damping on a time scale of few precession cycles \citep{l2006} 
would not contradict observations of RX J0720.4-3125. 
 
The glitch in RX J0720.4-3125 reported by \cite{vketal2007} by its
consequencies is similar to the one observed in an anomalous X-ray pulsar 
(AXP) 
CXOU J164710.2-455216 \citep{ietal2007,metal2007}. After the glitch
the luminosity of the source was increased, and its spectrum changed.
So, the jump in properties of the spectrum and luminosity of RX J0720.4-3125
proposed by \cite{vketal2007} can be directly related to a glitch, which is weaker
than in the case of CXOU J164710.2-455216 (still similarities in behavior of these
sources can be considered as a kind of support to the hypothesis of a
link between AXPs and XDINS). But the evolution of the NS parameters after the ``jump'' 
requires precession.

 Note, that the timing solution before MJD 52821, when a possible "glitch"
happened according to \cite{vketal2007}, 
can be relatively well described by the so-called cubic solution (the second
derivative of $\nu$ is non-zero), see \cite{vkk2007}. Spectral changes before
this date are not very large \citep{hetal2006, vkk2007}. After MJD 52821 the
timing solution is well described by a periodic function, see \cite{vkk2007}
(these authors studied several models with and without a glitch in their two papers),
and spectral changes follow this law, too
\citep{hetal2006}\footnote{However, van Kerkwijk and Kaplan (2007) propose
that the period is not close to $\sim$7 years, but is $\sim$4.3 years. Still, on a
short time scale -- since MJD 52821 -- this is not very certain.}. 
Based on that, I suggest that
the timing residuals might be also explained by a model without (or with
small) precession before the glitch, and strong precession after.
However, this particular model has never been tested
quantitatively against observational data. 

Glitches naturally can produce thermal afterglows \citep{hetal1997}. 
About 
$10^{38}$~--~$10^{43}$~ergs can be released in a glitch 
(in the case of RX J0720.4-3125 according to estimates of the 
increase in spin frequency by \citealt{vketal2007} this value is closer to $10^{38}$~erg).  
However, Hirano et al. showed that a thermal response of a NS to a glitch cannot 
produce a smooth 
temperature increase on the time scale of years. 
If surface temperature is
 increased just by few percent, 
as it is required by \cite{vketal2007}, then 
the brightening lasts just for few days 
(this corresponds to weak energy release). 
If we require a temperature 
rise for a long time, then the effect is
too strong \citep{hetal1997}. 
So, I conclude that spectral changes on a long time scale 
should be attributed to precession of the NS.

Glitches of AXPs (and soft gamma-repeaters) can be different  in nature with 
respect to radio pulsar glitches, as the former can be related to crust fracture
due to superstrong magnetic field. Still, the origin of a glitch is not important for
our discussion here.
``Normal'' glitches are quite common for long period pulsars, 
for example, PSR J1814-1744 with spin period about 4 seconds demonstrated 
a glitch \citep{js2006}.   So, RX J0720.4-3125 can glitch not only via 
the mechanism operating in magnetars, but also due to convenient mechanisms proposed
for normal radio pulsars. 
For them one can estimate the reccurence time following \cite{ab1994}.

If the glitch of RX J0720.4-3125 is due to unpinning, 
then using standard formulae
 \citep{ab1994} one obtains that
the reccurence time between two succesive glitches is:

\begin{equation}
t_{\mathrm{g}}=\frac{\delta \Omega}{\dot \Omega}.
\end{equation}
Parameter $\delta \Omega$ is the critical value of the difference between
the rotation frequencies of normal matter and the superfluid at a boundary
layer.
$\delta \Omega$ itself can be estimated as (Alpar \& Baykal 1994):

\begin{equation}
\delta \Omega = \Delta \Omega \frac{I}{I_{\mathrm{p}}},
\end{equation}
here $I_{\mathrm{p}}$ is the effective moment of inertia of the region of
a pinning layer.

Combining these two formulae one obtains the relation for 
the time between glitches:

\begin{equation}
t_{\mathrm{g}} =
\frac{2\, I}{I_{\mathrm{p}}} \frac{\Delta \Omega}{\Omega} t \propto t,
\label{eq2}
\end{equation}
where $t$ is the age of a pulsar, 
$t=\Omega/2\, \dot \Omega$.

Surprisingly, the time is about 10 years for RX J0720.4-3125. 
I.e. it is quite probable to
observe one since the discovery of this object. Then, one can expect to see similar 
phenomenae in other Magnificent seven objects. 
However, they are less studied, and may be 
 some  glitches are missed.

Still, to produce  luminosity and spectral changes, and to generate
Tkachenko waves, it is probably more natural to have a glitch due to a starquake. 
In the case of the quake model \citep{ab1994}
 the time between glitches is longer, 
about 300 years for RX J0720.4-3125. 

\begin{equation}
t_{\mathrm{g}} = \frac{2(A+B)\phi \Delta \Omega / \Omega}
{I_0\Omega \dot \Omega}.
\end{equation}
The estimate above was obtained assuming standard values \citep{ab1994} 
$A=10^{52}$ erg, $\phi=10^{-3}$,
$B=10^{48}$~erg,  and $I\sim 10^{45}$~g~cm$^2$

Then, we can be just lucky to find a glitch 
in $\sim 10$ years of
observations (but note, that it is not the only XDINS observed). 
Or, in 
XDINS quakes do not follow the formula for radio pulsars.

\cite{vketal2007} proposed that an accretion episode can be responsible for spectral
changes after a ``glitch'' in RX J0720.4-3125.
I think that this
is not a very probable reason for the origin of
the glitch and corresponding changes.  It is hardly possible to imagine that
if we observe such an episode just after 10 years of observations, other 
episodes were not frequent during the evolution of this source. With
frequent episodes of accretion of light elements a NS should follow a
slightly different cooling track \citep{ketal2006}. Such stars with
accreted envelopes are hotter in their youth, but colder after they mature. 
This is more similar to the properties of 1E1207.4-5209 and Kes 79
\citep{gh2007}. 

A remarkable difference between some NSs in supernova remnants
 (so-called CCOs) 
and XDINS in the solar vicinity can be due to the existence
 of accreted envelopes 
in the former. Note, that we do not observe
 descendants of 1E1207-like sources 
in our proximity. If they follow a
 standard cooling curve, like RX J0720.4-3125 
and other XDINS, then they have
 to be observed. On the other hand, we do not 
see ancestors of XDINS in
 supernova remnants. This can be related to the fact that CCOs descendants
are too cold at the age of XDINS to be easily detected by ROSAT, and vice versa
ancestors of XDINS are not hot enough to be easily found in some supernova remnants.

 In this note I neglect (as most, if not all, other authors who studied Tkachenko waves
in NSs) the influence of interaction between neutron vortex lines and magnetic flux tubes.
This interaction can significantly affect the velocity of waves, and so their period, and
to damp them.\footnote{This was noted to us by Prof. M. Ruderman.} 
The velocity of a Tkachenko wave can be estimated as:

\begin{equation}
V_\mathrm{T}= 1/2 (h \Omega/2\pi m_\mathrm{pair})^{1/2}= (\Omega k/8 \pi)^{1/2} \sim b/P.
\end{equation}
Here $m_\mathrm{pair}=2 m_\mathrm{n}$, and $b$ is the distance between vortex lines.
So, $P_\mathrm{T}\sim (R/b) P$ in the simple case when there are no interactions
with flux tubes or other complications. 
When vortices are able to ``communicate'' with the help of numerous 
magnetic flux tubes, the velocity of the wave can be larger, so the period of 
Tkachenko-like wave would be shorter. This question should be explored.

To conclude, in this short note I proposed that long period precession in
RX J0720.4-3125 can be related to Tkachenko waves, generated in a recent
glitch. 
In the proposed model before a glitch precession could be negligible.
The critical condition for the free precession excitation
is the  equality between Tkachenko wave
period and the period of free precession.

\acknowledgments
It is a pleasure to thank D.I. Jones, A.D. Kaminker,
M. Ruderman, R. Turolla and many colleagues from SAI for discussions.
I thank the referee for useful criticizm.
This work was partially supported by the RFBR grant  06-02-16025
and by the INTAS Foundation.


\end{document}